\documentclass[aps,prb,preprint,showpacs,groupedaddress]{revtex4} 

\usepackage{graphicx}

\begin{document}

\title{
Electron  and  spin  correlations in semiconductor
heterostructures: Quantum Singwi-Tosi-Land-Sj\"{o}lander theory }

\author{Nguyen Thanh Son}
\affiliation{ Center for Theoretical Physics, Massachusetts
Institute of Technology, 77 Massachusetts Avenue , Cambridge, MA
02139-4307, USA }

\author{Nguyen Quoc Khanh}
\email[]{nqkhanh@hcmuns.edu.vn}
\affiliation{ Department of Theoretical Physics, National
University in Ho Chi Minh City, 227-Nguyen Van Cu Str., District
5, Ho Chi Minh City, Vietnam }

\date{\today}

\begin{abstract}
We apply the quantum Singwi-Tosi-Land-Sj\"{o}lander (QSTLS) theory
for a study of many-body effects in the quasi-two-dimensional
(Q2D) electron liquid (EL) in \linebreak $GaAs/Al_{x}Ga_{1-x}As$
heterojunctions. The effect of the layer thickness is included
through a variational approach. We have calculated the density,
spin-density static structure factors, spin-dependent pair
distribution functions (PDF) and compared our results with those
of two-dimensional (2D) EL given in earlier papers. Using the
static structure factors (SSF) we have calculated various dynamic
correlation functions such as spin-dependent local-field factors
(LFF) and effective potentials of the Q2D EL. We have also
calculated the inverse static dielectric function of the 2D and
Q2D EL using different approximations. We find that the effect of
finite thickness on the dielectric function is remarkable and at
the intermediate values of wave number $q $ there is a significant
difference between the QSTLS and STLS results.
\end{abstract}

\pacs{73.21.Fg, 71.45.Gm, 71.10.-w}

\maketitle

\section{Introduction}

During the last decades quasi-two-dimensional electron systems
have been of considerable interest because of technological
relevance to high-mobility electronic devices. Many authors have
investigated Q2D systems applying various methods \cite{AFS} and
most of the theoretical calculations have been performed in the
framework of the random phase approximation (RPA). However, it is
well-known that the RPA neglects the Pauli and Coulomb hole
surrounding each particle and is not satisfactory approximation at
low densities \cite{MH}. Therefore a number of authors have
studied beyond-RPA effects by including the static LFF such as in
the Singwi-Tosi-Land-Sj\"{o}lander (STLS) approach \cite{STLS68}.
A common viewpoint is the need to incorporate dynamic correlations
and the importance of dynamic LFF is evidence. The dynamic LFF has
been introduced in several works via different schemes \cite{ST}.
One of these works is that of Hasegawa and Shimizu using the
Wigner distribution function \cite{HS}. The approach is usually
named as the quantum STLS and has been applied to 2D EL by Moudgil
and coworkers \cite{MAP}. Their calculations have been generalized
in our recent work \cite{KS} to the more realistic case of
electrons in Q2D semiconducting heterostructures investigated by
Bulutay and Tomak (BT) \cite{BT}.  The aim of this work is to
extend our previous investigations to include both electron and
spin correlations.

The outline of this paper is as follows. In Sec. II, we discuss
briefly the theoretical formalism of the self-consistent QSTLS
equations applied to the Q2D EL in GaAs/Al$_{x}$Ga$_{1 - x}$As
heterojunctions based on the variational approach proposed by
Bastard \cite{BA}. In Sec. III the results and discussion are
given, and in Sec. IV we summarize the main results and gather our
assessment of the performance of the QSTLS for semiconductor
heterostructures.
\section{Theory}

We consider a GaAs/Al$_{x}$Ga$_{1 - x}$As heterojunction where the
electrons move into the GaAs side and form 2D subbands. For an
accurate account of the electronic distribution we use Bastard's
variational approach given in BT's work, where the electron
effective masses $m_{A,B}$ and dielectric constants
\textit{$\varepsilon $}$_{A,B}$ are considered to be different in
the GaAs and Al$_{x}$Ga$_{1 - x}$As layers. In the regime where
only the lowest subband is populated the self-consistent equations
for density and spin-density static structure factors at
zero-temperature can be written as \cite{MAP} $$ \label{eq1} \chi
^d\left({q,\omega } \right) = \frac{\chi _0 \left({q,\omega }
\right)}{1 - V\left(q \right)\left[ {1 - G\left({q,\omega }
\right)} \right]\chi _0 \left({q,\omega } \right)} \eqno{(1a)} $$
$$ \label{eq2} S\left(q \right) = - \frac{2\hbar
}{n}\int\limits_0^\infty {Im\left[ {\chi ^d\left({q,\omega }
\right)} \right]d\omega } \eqno{(2a)} $$ $$ \label{eq3}
G\left({q,\omega } \right) = - \frac{1}{n}\int
{\frac{V(q')}{V(q)}\frac{\chi _0 \left({\vec {q},\vec
{{q}'};\omega } \right)}{\chi _0 \left({\vec {q},\omega }
\right)}\left[ {S\left({\left| {\vec {q} - \vec {{q}'}} \right|}
\right) - 1} \right]\frac{d\vec {{q}'}}{\left({2\pi } \right)^2}}
\eqno{(3a)} $$ $$ \label{eq4} \chi ^S\left({q,\omega } \right) = -
g^2\mu _B ^2\frac{\chi _0 \left( {q,\omega } \right)}{1 - V\left(q
\right)J(q,\omega)\chi _0 \left( {q,\omega } \right)} \eqno{(1b)}
$$ $$ \label{eq5} \tilde {S}\left(q \right) = \frac{\hbar }{2\pi
ng^2\mu _B ^2}\int\limits_{ - \infty }^\infty {Im\left[ {\chi
^S\left({q,\omega } \right)} \right]d\omega } \eqno{(2b)} $$ $$
\label{eq6} J\left({q,\omega } \right) =  \frac{1}{n}\int
{\frac{V(q')}{V(q)}\frac{\chi _0 \left({\vec {q},\vec
{{q}'};\omega } \right)}{\chi _0 \left({\vec {q},\omega }
\right)}\left[ {\tilde {S}\left( {\left| {\vec {q} - \vec {{q}'}}
\right|} \right) - 1} \right]\frac{d\vec {{q}'}}{\left({2\pi }
\right)^2}} \eqno{(3b)} $$ where $$ \label{eq7} \chi _0
\left({\vec {q},\vec {{q}'};\omega } \right) = - \frac{2}{\hbar
}\int {\frac{d\vec {p}}{h^2}\frac{f^0\left({\vec {p} +
\raise0.7ex\hbox{${\hbar \vec {{q}'}}$} \!\mathord{\left/
{\vphantom {{\hbar \vec {{q}'}}
2}}\right.\kern-\nulldelimiterspace}\!\lower0.7ex\hbox{$2$}}
\right) - f^0\left({\vec {p} - \raise0.7ex\hbox{${\hbar \vec
{{q}'}}$} \!\mathord{\left/ {\vphantom {{\hbar \vec {{q}'}}
2}}\right.\kern-\nulldelimiterspace}\!\lower0.7ex\hbox{$2$}}
\right)}{\omega - \raise0.7ex\hbox{${\vec {q}\vec {p}}$}
\!\mathord{\left/ {\vphantom {{\vec {q}\vec {p}}
m}}\right.\kern-\nulldelimiterspace}\!\lower0.7ex\hbox{$m$} + i0}}
\eqno{(4)} $$ with $f^0\left(\vec {p} \right)$ is the Fermi --
Dirac distribution function, $\chi _0 \left({\vec {q};\omega }
\right) = \chi _0 \left({\vec {q},\vec {q};\omega } \right)$ is
the free-electron response function and $$ \label{eq8} V\left(q
\right) = \frac{2\pi e^2}{q\bar {\varepsilon }}F\left(q \right)
\eqno{(5)} $$ is the Fourier transform of the effective potential.
Here $F(q)$ is the form factor to the Coulomb interaction due to
the layer thickness given in the BT's work and $\bar {\varepsilon
}$= (\textit{$\varepsilon $}$_{A}$\textit{+$\varepsilon
$}$_{B})/2$ is the average dielectric constant. A strictly 2D
electron gas with $\delta $-function density distribution is
obtained by setting $F(q)$ = 1 in Eq. (5)$.$ To solve the sets of
Eqs. (1a-3a) and (1b-3b) we have used the procedure proposed by de
Freitas \textit{et al. }\cite{FIS} and obtained the following
self-consistent equations $$ \label{eq9} S\left(q \right) =
\frac{q^2k_F }{\pi ^2n}\int\limits_0^{\gamma \left(q \right)} f\
\left(q,\beta  \right)d\beta  \eqno{(6a)} $$ $$ \label{eq10}
G\left({q,\beta } \right) = - \frac{1}{\left({2\pi }
\right)^2n}\int\limits_0^\infty {dq'\int\limits_0^{2\pi } {d\phi
\mbox{ }h\left({\vec {q},\vec {q}';\beta } \right)\left[
{S\left({q'} \right) - 1} \right]} } \eqno{(6b)} $$

$$ \label{eq11} \tilde {S}\left(q \right) = \frac{q^2k_F }{\pi
^2n}\int\limits_0^{\gamma \left(q \right)} {\tilde
{f}\left({q,\beta } \right)d\beta } \eqno{(7a)} $$

$$ \label{eq12} J\left({q,\beta } \right) = \frac{1}{\left({2\pi }
\right)^2n}\int\limits_0^\infty {d{q}'\int\limits_0^{2\pi } {d\phi
\mbox{ }h\left({\vec {q},\vec {{q}'};\beta } \right)\left[ {\tilde
{S}\left({q}' \right) - 1} \right]} } \eqno{(7b)} $$ with $$
\label{eq13} f\left(q,\beta  \right) = \left({\sqrt {1 -
\frac{q^2\sin ^2\beta }{4k_F^2 }} + \frac{\cot ^2\beta }{\sqrt {1
- \frac{q^2\sin ^2\beta }{4k_F^2 }} }} \right)\frac{\left({1 -
\cos \beta } \right)}{q + \frac{m}{\pi \hbar ^2}qV\left(q
\right)\left[ {1 - G\left({q,\beta } \right)} \right]\left( {1 -
\cos \beta } \right)}, \eqno{(8a)} $$ $$ \label{eq14} \tilde
{f}\left({q,\beta } \right) = \left({\sqrt {1 - \frac{q^2\sin
^2\beta }{4k_F^2 }} + \frac{\cot ^2\beta }{\sqrt {1 -
\frac{q^2\sin ^2\beta }{4k_F^2 }} }} \right)\frac{\left({1 - \cos
\beta } \right)}{q + \frac{m}{\pi \hbar ^2}qV\left(q
\right)J\left({q,\beta } \right)\left({1 - \cos \beta } \right)}
\quad , \eqno{(8b)} $$ $$ \label{eq15} \gamma \left(\rho \right) =
\left\{ {\begin{array}{l}
 \pi \mathord{\left/ {\vphantom {\pi 2}} \right. \kern-\nulldelimiterspace}
2\mbox{ },q \le 2k_F \\
 \mbox{sin}^{-1}\left({\raise0.7ex\hbox{${2k_F }$} \!\mathord{\left/
{\vphantom {{2k_F }
q}}\right.\kern-\nulldelimiterspace}\!\lower0.7ex\hbox{$q$}}
\right)\mbox{ },q > 2k_F \\
 \end{array}} \right.
\eqno{(9)} $$ $$ \label{eq16} h\left({\vec {q},\vec {q}';\beta }
\right) = \frac{q'\left[ {1 - r\left( {\vec {q},\vec {q} - \vec
{q}';\beta } \right)} \right]\left({q - q'\cos \phi } \right)}{q(1
- \cos \beta)}\frac{V\left({\sqrt {q^2 + q'^2 - 2qq'\cos \phi } }
\right)}{V\left(q \right)} \eqno{(10)} $$ where $r(\vec {q},\vec
{q} - \vec {q}';\beta)$  is the positive root of the following
equation $$ \label{eq17} \left[ {\frac{\vec {q}.\left({\vec {q} -
\vec {q}'} \right)}{\vec {q}^2}} \right]^2 = \frac{1}{1 -
r^2}\frac{4k_F^2 }{\vec {q}^2} - \frac{1}{r^2}\left({\frac{4k_F^2
}{q^2\sin ^2\beta } - 1} \right)\cos ^2\beta \quad. \eqno{(11)} $$
Here $n $, $k_F = \sqrt {2\pi {\kern 1pt} n} $ and \textit{$\phi
$} are the electron density, Fermi wave number and angle between
$\vec {q}$ and $\vec {q}'$, respectively \cite{MAP}.  Using the
integration by parts we can write the density and spin-density
static structure factors as $$ \label{eq18} S\left(q \right) =
\frac{2q^2}{\pi}\int\limits_0^{\gamma \left(q \right)} {\cot \beta
\sqrt {1 - \frac{q^2\sin ^2\beta }{4}} \frac{q\sin \beta + \sqrt 2
r_S F\left(q \right)(1 - \cos \beta)^2\frac{\partial
G\left({q,\beta } \right)}{\partial \beta }}{\left({q + \sqrt 2
r_S F\left(q \right)\left[ {1 - G\left({q,\beta } \right)}
\right]\left({1 - \cos \beta } \right)} \right)^2}} d\beta \quad ,
\eqno{(12a)} $$ $$ \label{eq19} \tilde {S}\left(q \right) =
\frac{2q^2}{\pi}\int\limits_0^{\gamma \left(q \right)} {\cot \beta
\sqrt {1 - \frac{q^2\sin ^2\beta }{4}} \frac{q\sin \beta - \sqrt 2
r_S F\left(q \right)(1 - \cos \beta)^2\frac{\partial
J\left({q,\beta } \right)}{\partial \beta }}{\left({q + \sqrt 2
r_S F\left(q \right)J\left({q,\beta } \right)\left({1 - \cos \beta
} \right)} \right)^2}} d\beta \eqno{(12b)} $$ These forms of the
structure factors are appropriate for the numerical integration
because the integrands remain finite in the limit $\beta \to 0$.

\section{Results and discussions}

In our recent paper \cite{KS} we have solved the equations (6a-7a)
for 2D and Q2D electron liquids in semiconductor heterojunctions.
In the case of 2D EL we have obtained the SSF and PDF in good
agreement with Monte Carlo results. In this paper we solve the
equations (6b-7b) for the Q2D electron liquid in
GaAs/Al$_{x}$Ga$_{1 - x}$As heterojunctions having a step-barrier
potential with $U_{b} = $0.3eV , \textit{$\varepsilon $}$_{A} =
$13,\textit{ $\varepsilon $}$_{B}$ = 12.1, $m_{A} =$0.07$m_{e}$
and $m_{B} = $0.088$m_{e}$ , where $m_{e}$ is the vacuum mass of
the electron \cite{BT}.  Using the obtained results we calculate
the spin-dependent pair-correlation functions, dynamic local-field
factors, spin-dependent effective potentials, inverse static
dielectric function and compare our results with those given in
Refs.7 and 8.
 \subsection{Static spin-density structure
factors }

The static structure factor $S(q)$ of the 2D EL was shown in Fig.
1 of our recent paper \cite{KS}.  In this work we have calculated
the spin-density SSF $\tilde {S}(q)$ of 2D and Q2D EL by solving
the Eqs. (6b) and (7b) in the self-consistent way for several
values of electron density and the results are shown in Fig. 1. It
is seen that in the case of 2D EL our results are similar to those
of Ref. 6 and the effect of the layer thickness is remarkable for
a wide range of electron densities.

\subsection{ Spin-dependent electron pair-correlation functions }

The spin-symmetric and spin-antisymmetric electron
pair-correlation functions can be expressed as $$ \label{eq20} g_{
\uparrow \uparrow } \left(r \right) = 0.5\;\left[ {g\left(r
\right) + \tilde {g}\left(r \right)} \right] \eqno{(13)} $$ $$
\label{eq21} g_{ \uparrow \downarrow } \left(r \right) =
0.5\;\left[ {g\left(r \right) - \tilde {g}\left(r \right)} \right]
\eqno{(14)} $$ where $$ \label{eq22} g\left(r \right) = 1 +
\int\limits_0^\infty {qJ_0 \left({qr} \right)\left[ {S\left(q
\right) - 1} \right]dq} \eqno{(15a)} $$ $$ \label{eq23} \tilde
{g}\left(r \right) = \int\limits_0^\infty {qJ_0 \left({qr}
\right)\left[ {\tilde {S}\left(q \right) - 1} \right]dq}
\eqno{(15b)} $$

To study the effect of electron density we show in Fig. 2. the
spin-dependent PDF $g_{ \downarrow \uparrow } (r)$ of 2D EL for
different values of density parameter $r_{s}. $ We observe that
our PDFs differ from those of Moudgil and coworkers at small
values of $r $. This difference in the behavior of PDFs stems from
the incorrect results of Moudgil and coworkers for the SSF at
large $q's $ discussed in our previous paper \cite{KS}$.$

To study the effect of the layer thickness on PDFs we have
calculated the spin-dependent electron pair-correlation functions
$g_{ \downarrow \downarrow } (r)$ of 2D and Q2D EL for different
values of electron density parameter $r_{s}.  $ We find that $g_{
\downarrow \downarrow } (r)$ is almost independent of $r_{s}$  and
is therefore plotted in Fig. 3 only for $r_{s} = 3.$ We observe
from the figure that the effect of the layer thickness is
considerable for an intermediate region of the inter-particle
distance.

\subsection{ Spin-symmetric and spin-antisymmetric dynamic
local-field factors: }

To compare our results with those of Moudgil \textit{et al.} and
to study the effect of the layer thickness we show in Figs. 4 and
5 the spin-dependent dynamic LFFs of 2D and Q2D EL as a function
of $\omega $ for $q =$ 1.1$k_{F}$ and $r_{s} =$ 3.  We observe
that our results for 2D EL are similar to those given in Ref.6
and the layer thickness has a remarkable influence on
\textit{G(q,$\omega)$} and \textit{J(q,$\omega)$} for a wide range
of frequencies $\omega $.

\subsection{ Effective dynamic potentials}

Spin-symmetric and spin-antisymmetric dynamic effective potentials
can be calculated from the spin-symmetric and spin-antisymmetric
dynamic LFF as $$ \label{eq24} V_{eff}^s \left({q,\omega } \right)
= V\left(q \right)\left[ {1 - G\left( {q,\omega } \right)} \right]
\eqno{(16a)} $$ $$ \label{eq25} V_{eff}^a \left({q,\omega }
\right) = V\left(q \right)J(q,\omega) \eqno{(16b)} $$

The effective potentials of the 2D and Q2D EL obtained in the
static limit ( i.e., $\omega $ = 0) for $r_{s}$ = 1 and 3 are
shown Figs. 6 and 7. We observe remarkable differences in the
results of 2D and Q2D EL for all values of $q$ and $r_{s}. $ We
note that because of incorrect results for the SSF given in Ref. 6
our results for both spin-symmetric and spin-antisymmetric static
effective potentials of 2D EL differ considerably from those of
Moudgil and coworkers in the region of large values of $q \quad (q
> $3.5$q_{F}$).

Following the authors of Ref.6 we define the spin-dependent
effective dynamic potential as $$ \label{eq26} V_{eff \uparrow
\uparrow } = V_{eff}^s \left({q,\omega } \right) + V_{eff}^a
\left({q,\omega } \right) \eqno{(17)} $$ $$ \label{eq27} V_{eff
\uparrow \downarrow } = V_{eff}^s \left({q,\omega } \right) -
V_{eff}^a \left({q,\omega } \right) \eqno{(18)} $$

The real and imaginary spin-dependent effective dynamic potentials
of 2D and Q2D EL for $r_{s}$ = 1 and $q = $1.1$k_{F}$  are
plotted, respectively, in Figs. 8 and 9. It is seen from the
figures that the effect of layer thickness is significant for a
wide range of frequency $\omega $ and our results for 2D EL are
similar to those of Moudgil and co-workers \cite{MAP}. We note
that the values of the real spin-dependent effective dynamic
potentials shown in the Fig. 9(a) of Ref.6 is not correct.  By
using the values of $V_{eff}^s \left({q,\omega } \right)$ and
$V_{eff}^a \left({q,\omega } \right)$ at $\omega $ = 0 we can see
that the authors of Ref.6 have divided the correct values of
$V_{eff \uparrow \uparrow } (q,\omega)$ and $V_{eff \uparrow
\uparrow } (q,\omega)$ by 2.

\subsection{ Inverse static dielectric function }

Finally, we calculate the inverse static dielectric function of 2D
and Q2D EL for $r_{s} = $3 using different approximations. The
figure 10 shows that the QSTLS results differ less from those of
the STLS approximation in the Q2D EL than in the 2D EL. However,
at the intermediate values of $q $ there is a significant
difference between the QSTLS and STLS results.

\section{Conclusions}

Using the QSTLS approximation we have studied the electron and
spin correlations in GaAs/Al$_{x}$Ga$_{1 - x}$As heterojunctions.
We have calculated the spin-dependent PDF, SSF, dynamic LFF,
effective potential and inverse static dielectric function of 2D
and Q2D EL. We have shown that the results for Q2D EL in
semiconductor heterostructures differ remarkably from those of 2D
EL and the difference between the QSTLS and the STLS results of
Q2D EL is considerable.  It is hoped that our results will be of
help in investigating the effect of electron and spin correlations
on properties of Q2D electron systems.

\begin{acknowledgments}
We gratefully acknowledge the financial support from the National
Program for Basic Research.
\end{acknowledgments}

\newpage
\begin{figure}
\includegraphics{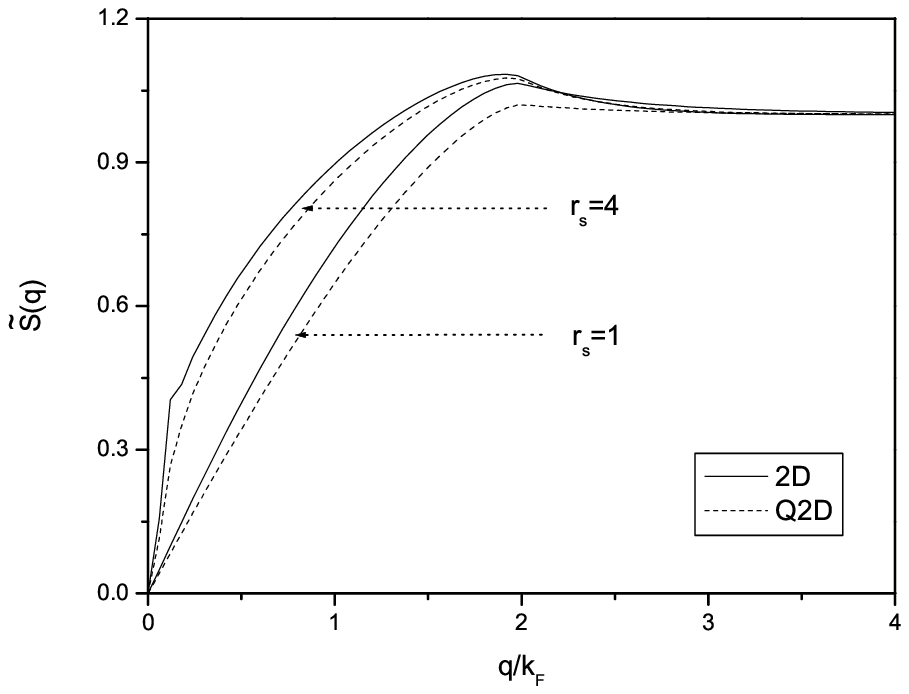} \caption{Static
spin-density structure factors for $r_{s} = $1, 2, 3 and 4.}
\end{figure}

\begin{figure}
\includegraphics{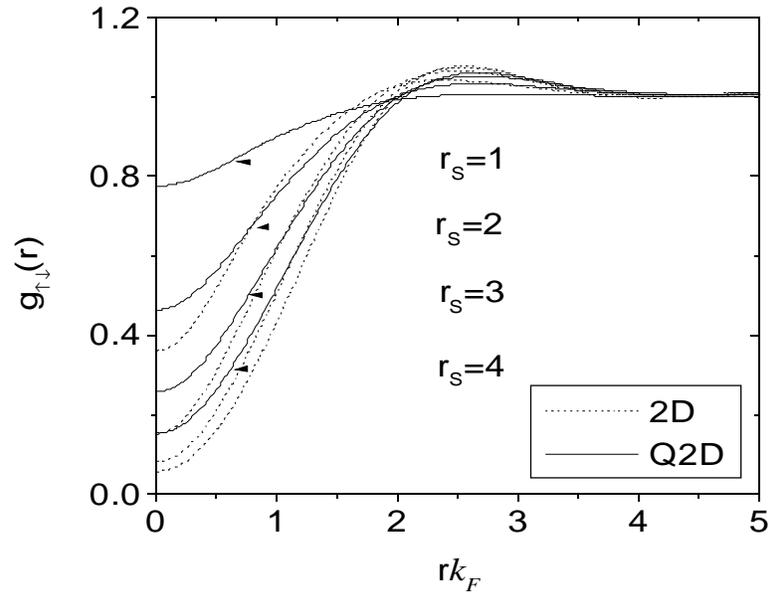}
\caption{Spin-dependent pair-correlation functions $g_{ \downarrow
\uparrow } (r)$ for $r_{s} = $1, 2, 3 and 4.}
\end{figure}
\begin{figure}
\includegraphics{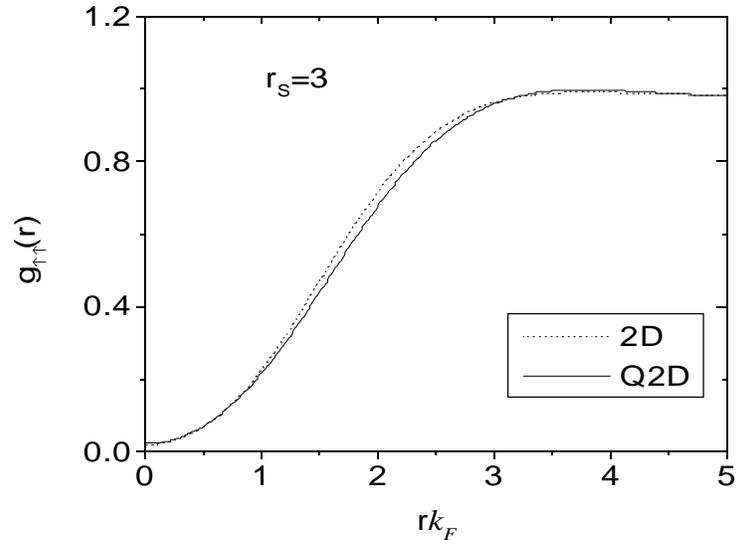}
\caption{Spin-dependent pair-correlation functions $g_{ \downarrow
\downarrow } (r)$ of the 2D and Q2D EL for $r_{s} = $3.}
\end{figure}
\begin{figure}
\includegraphics{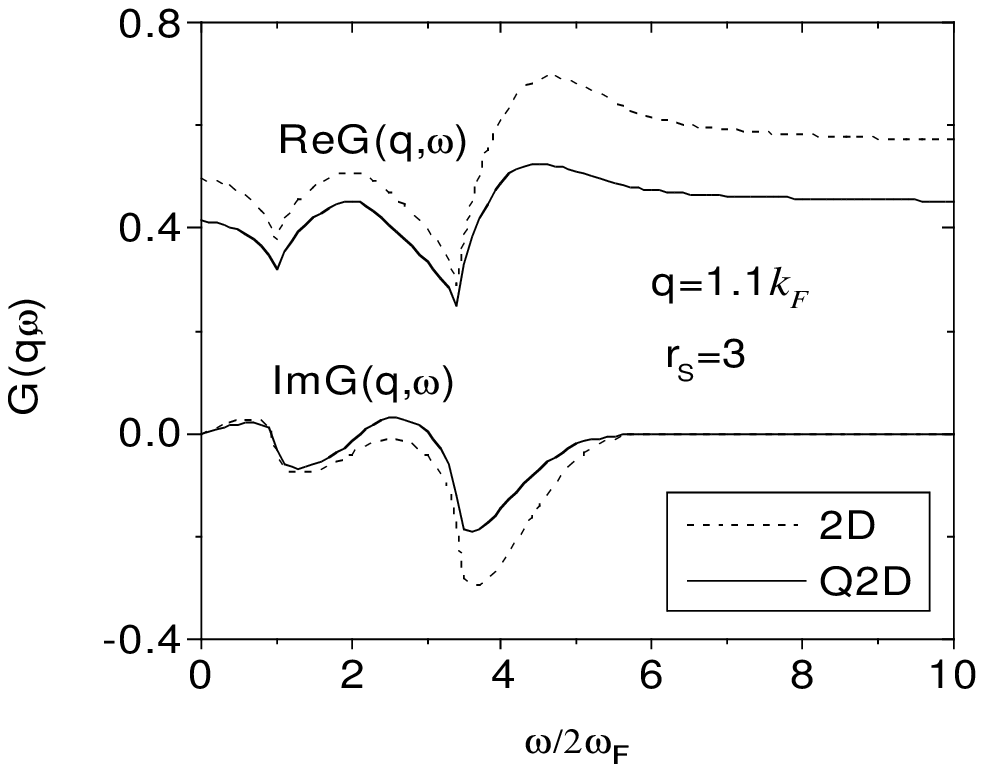}
\caption{Spin-symmetric dynamic structure factors for $r_{s}$ = 3
and $q = $1.1$k_{F}.$}
\end{figure}
\begin{figure}
\includegraphics{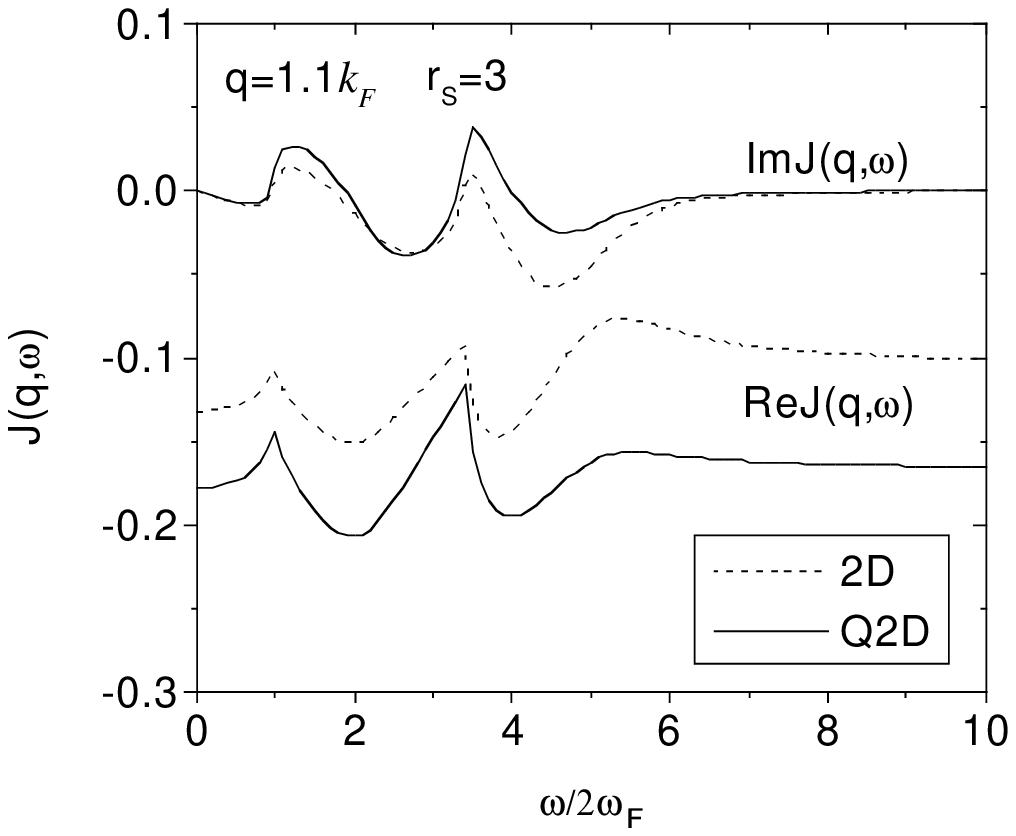}
\caption{Spin-antisymmetric dynamic structure factors for $r_{s}$
= 3 and $q = $1.1$k_{F}.$}
\end{figure}
\begin{figure}
\includegraphics{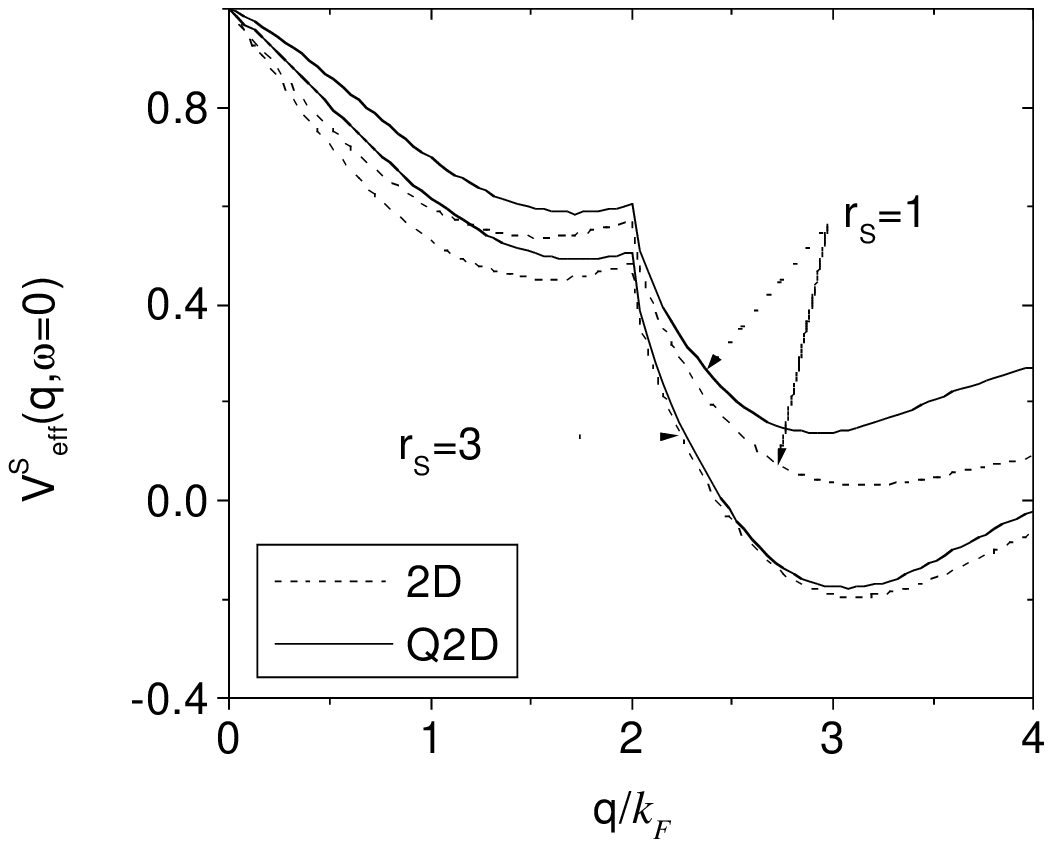}
\caption{Static spin-symmetric effective potentials for $r_{s}$ =
1 and 3.}
\end{figure}
\begin{figure}
\includegraphics{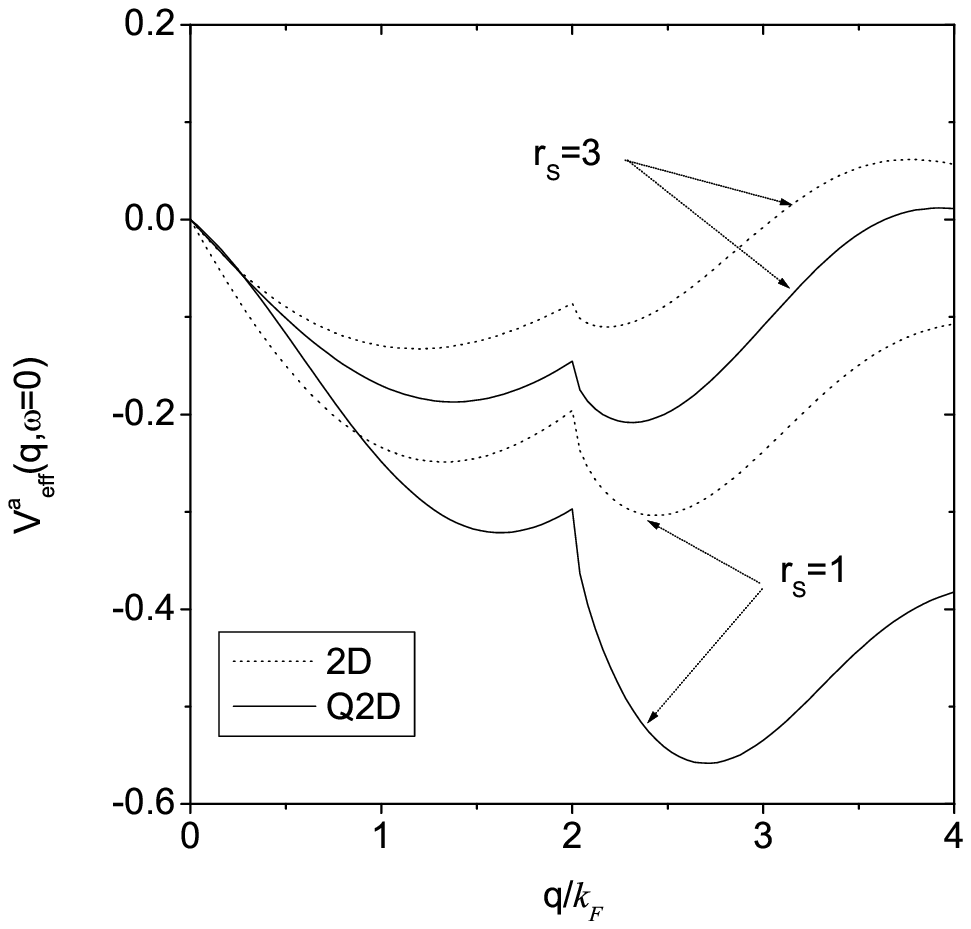}
\caption{Static spin-antisymmetric effective potentials for
$r_{s}$ = 1 and 3.}
\end{figure}
\begin{figure}
\includegraphics{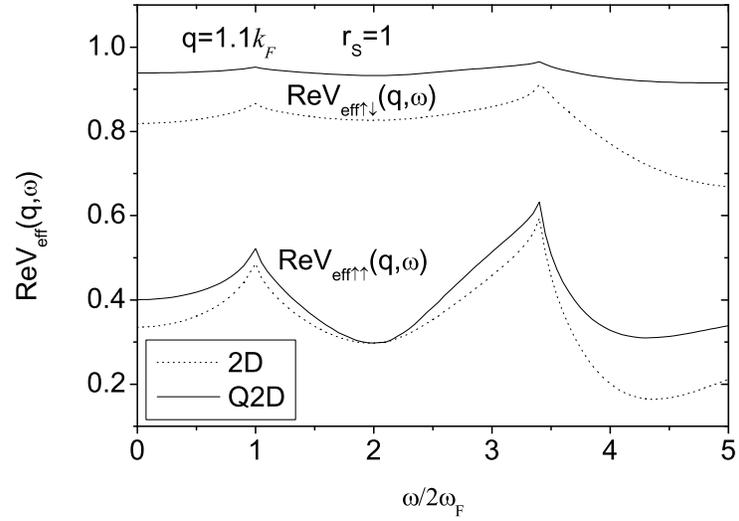}
\caption{Real spin-dependent effective dynamic potentials $r_{s}$
= 1 and $q = $1.1$k_{F}$.}
\end{figure}
\begin{figure}
\includegraphics{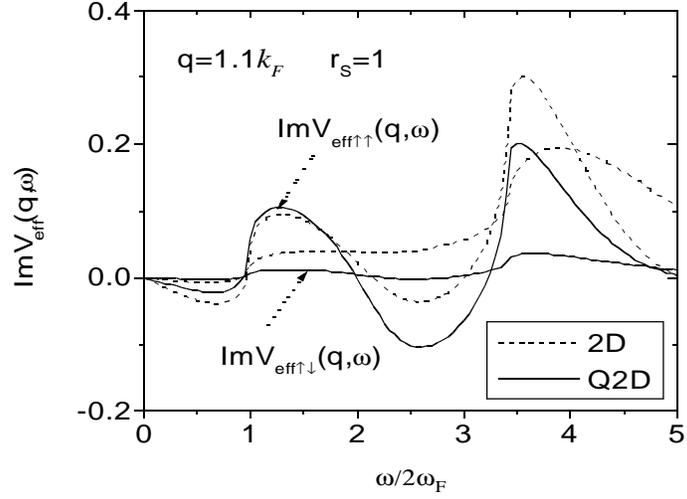}
\caption{Imaginary spin-dependent effective dynamic potentials
$r_{s}$ = 1 and $q = $1.1$k_{F}$.}
\end{figure}
\begin{figure}
\includegraphics{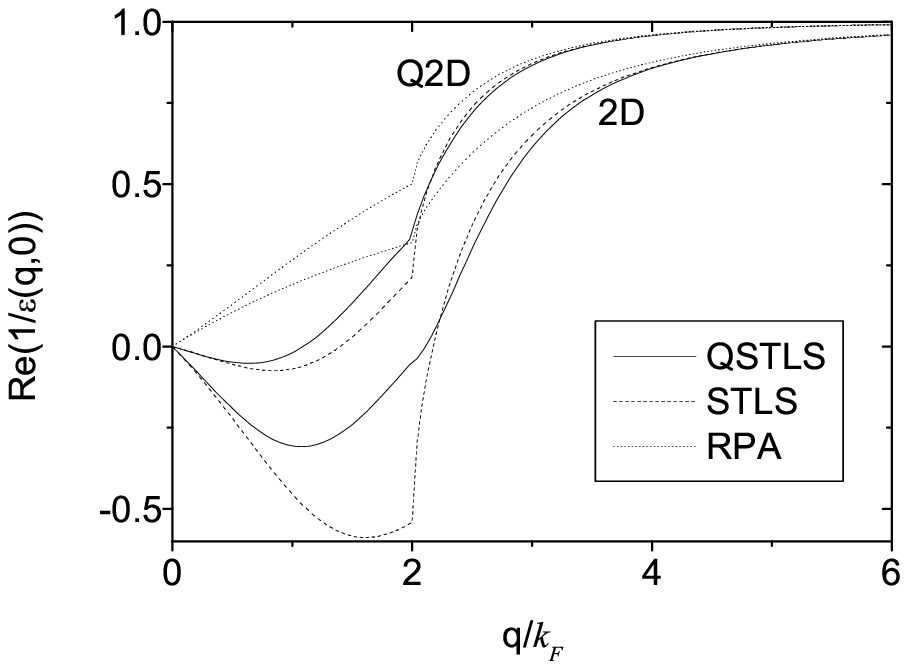}
\caption{Inverse static dielectric function of 2D and Q2D EL for
$r_{s} = $3.}
\end{figure}


\begin{thebibliography}{99}
\bibitem{AFS}
T. Ando, A.B. Fowler, and F. Stern, Rev. Mod. Phys. \textbf{54} ,
437 (1982).
\bibitem{MH}
G.D. Mahan, \textit{Many-Particle Physics,} 2$^{nd}$ ed. (Plenum
Press, New York , 1990).
\bibitem{STLS68}
K. S. Singwi, M. P. Tosi, R. H. Land , and A. Sj\"{o}lander,
Phys.Rev.\textbf{176} , 589 (1968).
\bibitem{ST}
K.S. Singwi and M.P. Tosi, in \textit{Solid State Physics}, edited by H. Ehrenreich, F. Seitz, and D.Turnbull
(Academic, New York , 1981), Vol. 36, p.177.
\bibitem{HS}
T.Hasegawa and M.Shimizu, J.Phys.Soc.Jpn. \textbf{38} , 965 (1975).
\bibitem{MAP}
R.K. Moudgil , P.K. Ahluwalia, and K.N. Pathak, Phys. Rev. B
\textbf{52} , 11945 (1995).
\bibitem{KS}
Nguyen Quoc Khanh and Nguyen Thanh Son, Physica B \textbf{344} , 176
(2004)
\bibitem{BT}
C.Bulutay and M.Tomak , Phys. Rev. B \textbf{54} , 14643 (1996).
\bibitem{BA} G. Bastard,
\textit{Wave Mechanics Applied to Semiconductor Heterostructures} (Les Editions de Physique, Les Ulis Cesex, 1988) p.155.
\bibitem{FIS}
U. de Freitas, L.C. Ioriatti , and N. Studart, J. Phys. C
\textbf{20} , 5983 (1987).



\end{thebibliography}
\end{document}